\documentclass[english]{revtex4-1}
\usepackage[T1]{fontenc}
\usepackage[latin9]{inputenc}
\usepackage{graphicx}

\makeatletter
 
 \@ifundefined{textcolor}{}
 {%
   \definecolor{BLACK}{gray}{0}
   \definecolor{WHITE}{gray}{1}
   \definecolor{RED}{rgb}{1,0,0}
   \definecolor{GREEN}{rgb}{0,1,0}
   \definecolor{BLUE}{rgb}{0,0,1}
   \definecolor{CYAN}{cmyk}{1,0,0,0}
   \definecolor{MAGENTA}{cmyk}{0,1,0,0}
   \definecolor{YELLOW}{cmyk}{0,0,1,0}
 }

\makeatother

\usepackage{babel}
\begin{document}

\title{A wire chamber for educational purposes}

\author{S. Gurbuz}

\address{Bogazici Univ., Physics and Astronomy Dept., Bebek, Istanbul, Turkey}

\author{G. Unel}

\address{University of California at Irvine, Physics and Astronomy Dept.,
Irvine, CA, USA}

\author{S. Erhan}

\address{University of California at Los Angeles, Physics and Astronomy Dept.,
Los Angeles, LA, USA}
\begin{abstract}
Gaseous detectors with sense wires are still in use today in small
experiments as well as modern ones as those at the LHC. This short
note is about the construction of a small wire chamber with limited
resources, which could be used both as an educational tool and also
as a tracker in small experiments. The particular detector type selected
for this work is the so called ``Delay Wire Chamber'': it has only
two output channels per plane and can be made fully gas tight for
educational operations. The design can be made with free software
tools, and the construction can be achieved by relatively simple means.
\end{abstract}
\maketitle

\section{Introduction}

\begin{figure}[h]
\begin{centering}
\includegraphics[width=0.8\textwidth]{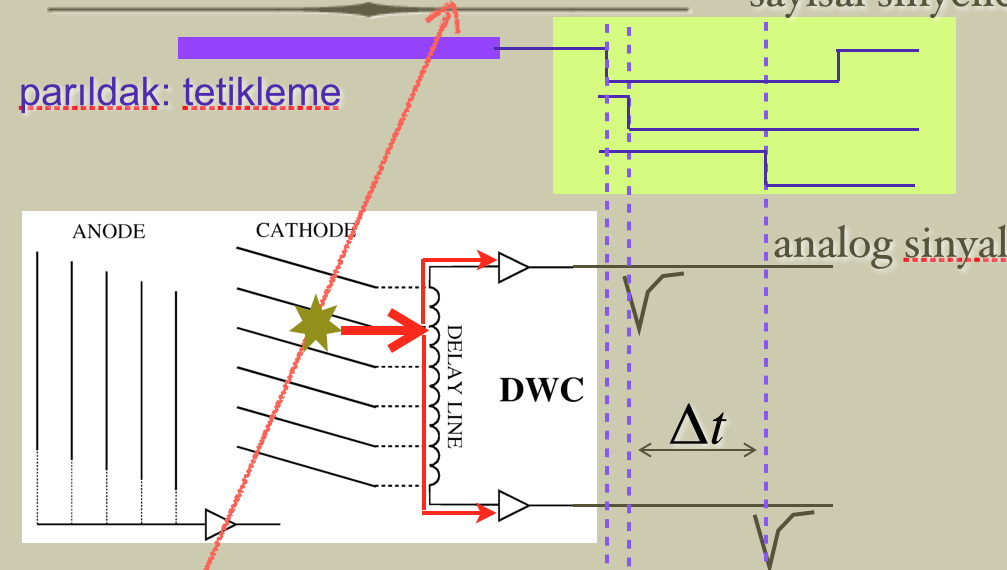}
\par\end{centering}

\protect\caption{A typical ``delay'' wire chamber\label{fig:dwc}}
\end{figure}

Theory of operation for the wire chambers has been studied extensively
in the past \cite{charpak-sauli}. As a general principle, these detectors
measure the drift time for the electrons from the ionized gas they
contain to the sense wires under the influence of a uniform electric
field generated by an applied high voltage. We describe here construction
of a so-called ``delay'' wire chamber (DWC) \cite{dwc-def}, a simplified
version of the well-known multi-wire proportional chamber. The first
change is on the cathode: instead of using a conductor plane, it is
made of a set of closely distanced wires (about 2mm pitch) placed
transversally to the anode wires. When the ionization electrons reach
the anode wires, an image charge is produced on the cathode wires.
An additional simplification is in the read-out: instead of acquiring
data from each individual wire, the signals are accumulated on a delay
line which is then read from both ends as shown in Fig. \ref{fig:dwc}.
The timing difference between the two signals ($\Delta t$) can be
converted into position information by a simple linear approximation:
$x=\alpha\times\Delta t+\beta$ where the calibration coefficients
$\alpha$ and $\beta$ are to be determined experimentally. Such a
simple detector has somewhat limited position precision, about 200
microns. However given its merits, such as the usage of non-flammable
gas (such as Ar-CO2 mixture) and simplicity of the readout circuitry,
such a compromise can be deemed acceptable for small experiments or
for educational purposes.

The prototype DWC described in this note is prepared for a simple
setup: it will be used in the magnetic spectrometer of the SPP proton
beamline \cite{spp}. Therefore the prototype is even further simplified
to measure position only in one plane.

\section{DWC Design}

\begin{figure}[h]
\begin{centering}
\includegraphics[width=0.3\textwidth]{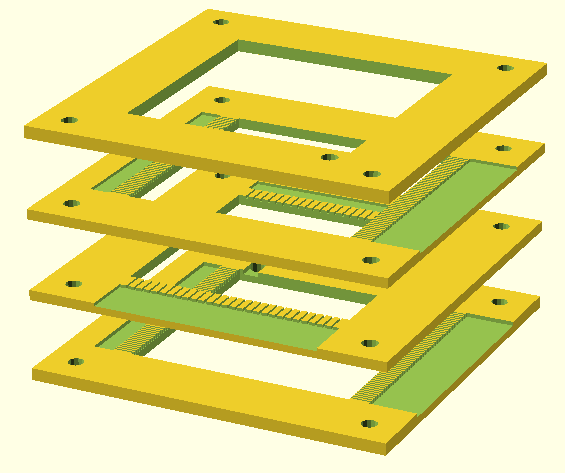} \includegraphics[width=0.5\textwidth]{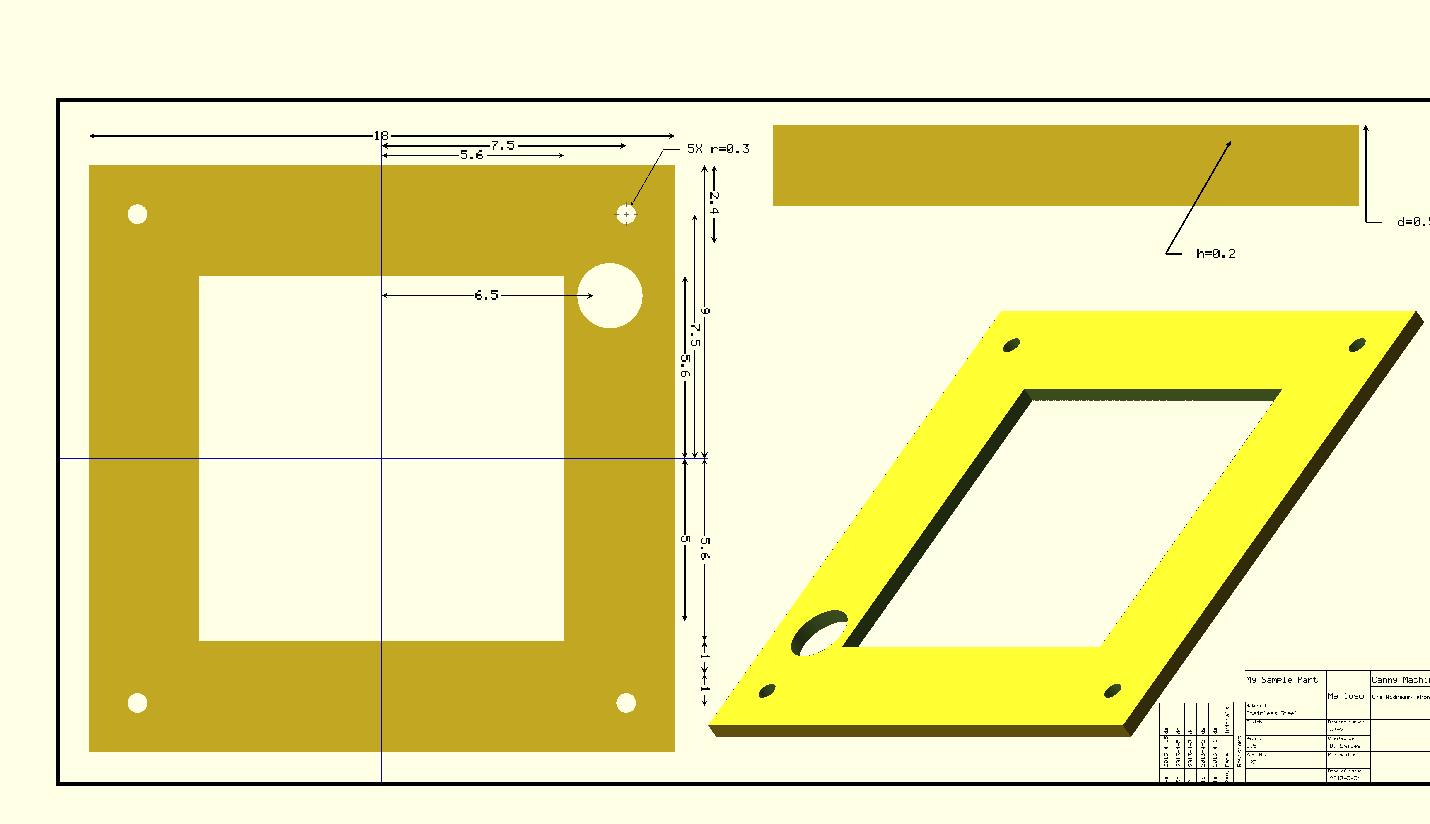}
\par\end{centering}

\protect\caption{Delay Wire Chamber 3D design on the left and engineering drawing on
the right.\label{fig:DWC-3D}}
\end{figure}

The structure to host the anode and cathode wires is made of glass-reinforced
epoxy laminate sheets that are also used in the printed circuit board
(PBC) making. To cope with the mechanical stress caused by the wires,
and to cope with the high voltages necessary for the electron drift,
PCB type flame retardant grade 4 (FR4) is selected \cite{FR4a,FR4b}.
FR4 has also the self-extinguishing and near zero water absorption
properties, making it ideal for gaseous detector construction. The
wire chamber design was made using software tools which are either
open source or free for educational use. The mechanical design for
the different layers of the DWC prototype is made using openSCAD,
available for free for a number of different operating systems \cite{scad}.
This software tool allows both a 3-D view of the product and also
standard engineering drawings. An example to both options can be seen
in Fig. \ref{fig:DWC-3D}. The individual FR4 layers, each having
5mm thickness, are designed with the goal of easy construction and
de-construction. 

\begin{figure}[h]
\begin{centering}
\includegraphics[bb=0bp 340bp 565bp 473bp,clip,width=0.8\textwidth]{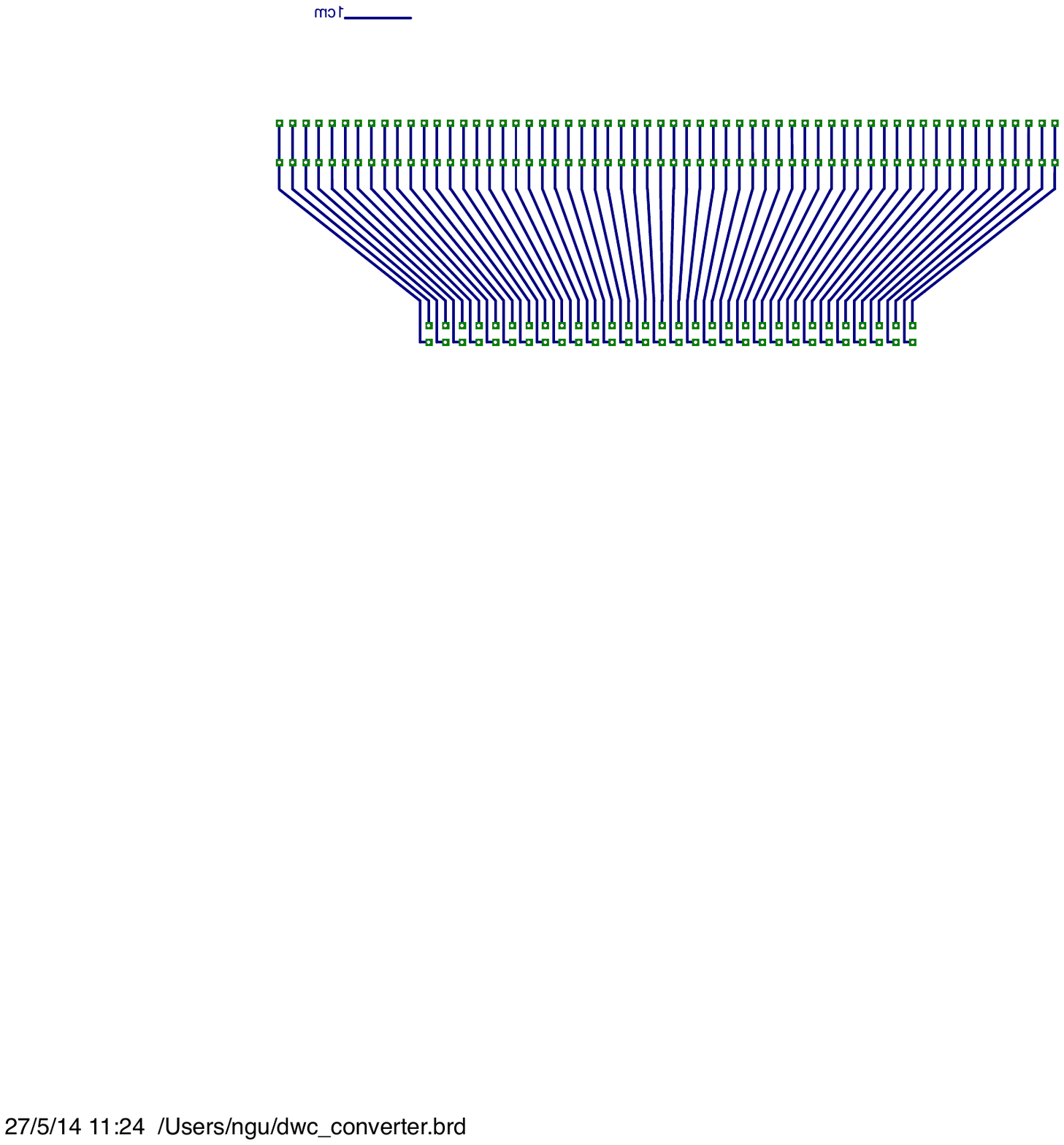} 
\par\end{centering}

\protect\caption{Pitch converter from 2mm to 0.1 inch.\label{fig:2mm254}}
\end{figure}

The top and bottom FR4 layers also contain the gas connectors to flush
the Ar-CO2 mixture. One could notice the narrow ditch that follows
the outer edge of all of the inner FR4 layers. This is the o-ring
housing to render the active part of the detector gas tight. The outer
faces of the top and the bottom inner openings should be covered with
a thin film to keep the Ar-CO2 gas mixture in and the water vapor
from the surroundings out while letting in as many of the incoming
particles. The topmost layer's inner face will house the first layer
of cathode wires. Following the literature, Copper wires with 2\%
beryllium content will be installed with 2mm pitch. The upper side
of the next layer houses the ditches to match the soldering joints
of the first cathode wires, and its lower side will house the anode
wires. These are selected to be Gold plated tungsten, with 4mm separation.
The third layer is a copy of the uppermost one, except it has ditches
on the upper side to match the soldering joints of the anode wires.
The last (bottommost) layer contains the usual soldering joint protection
ditches on its upper side. The ditches are 1mm deep to host PCB segments
with appropriate hole spacing, which is 2mm for both anode and cathode
planes. The connector side is made to extend out of the FR4 frame.
It is to host a specially designed PCB that will convert the 2mm pitch
on the wire side to 2.54mm (0.1 inch) on the connector side. The design
of this connector, made using EAGLE \cite{eagle} (free for educational
use) can be seen in Fig. \ref{fig:2mm254}.

The electronic readout circuit design with appropriate delays between
the anode wires has been around for more than 20 years. Unfortunately
some of the integrated circuits are not available anymore. A new circuit
design with smaller footprint and with more up-to-date circuits is
in progress, and will be used.

\section{Machining and Construction}

\begin{figure}[h]
\begin{centering}
\includegraphics[width=0.4\textwidth]{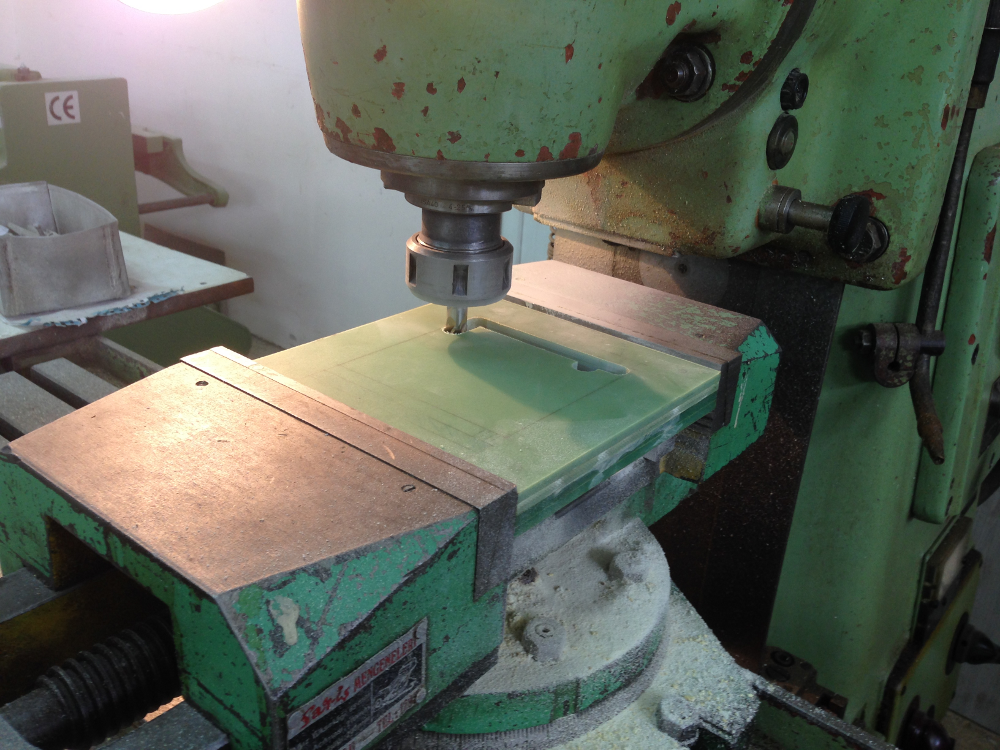} \includegraphics[width=0.4\textwidth]{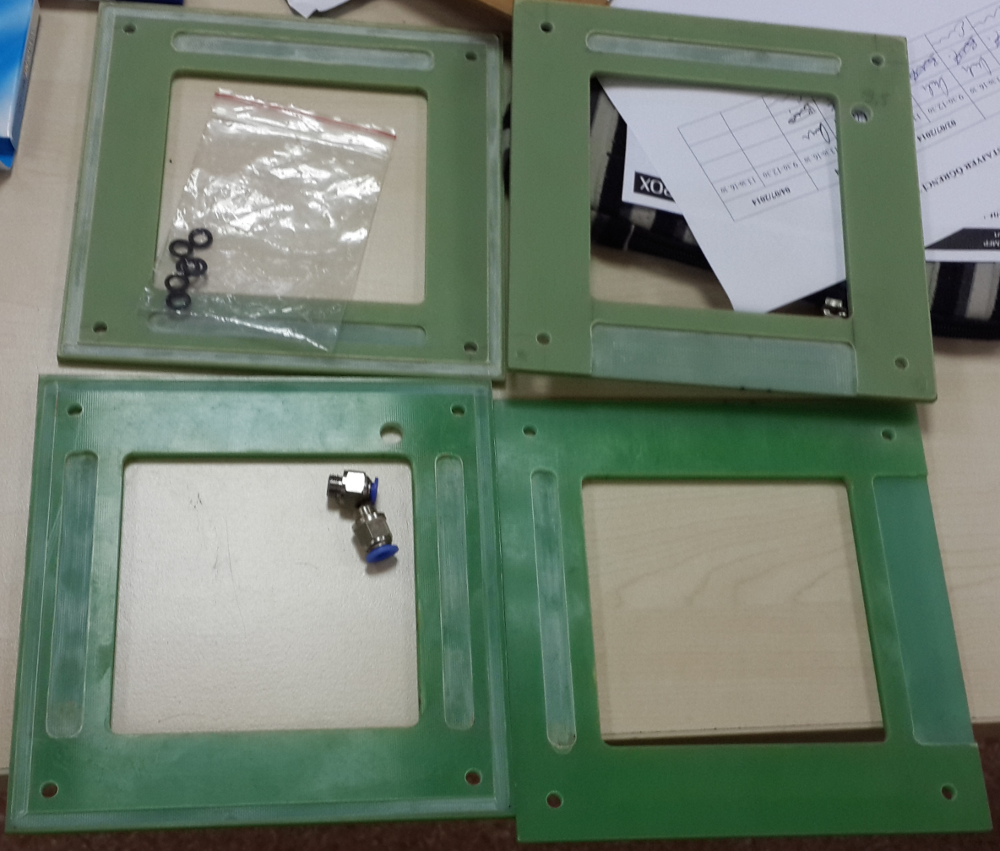}
\par\end{centering}

\protect\caption{Machining the FR4 with a drilling machine with simple controls\label{fig:Machining-the-FR4}}
\end{figure}

The FR4 frames were initially processed with a simple drilling machine with mechanical control wheels (which can be available even in high schools) , all four stacked together as in Fig. \ref{fig:Machining-the-FR4}
left side. After rendering of the common design features such as the
inner window and the main screw holes, each layer is machined individually
based on the design discussed in the previous section. The resulting
four planes can be seen on the same figure right side. The same photo
contains the gas connectors and the o-rings for the main screws. Note
that due to the simplicity of the tools, the ditches have rounded
corners to which the PCBs will have to be adapted. 

\begin{figure}[h]
\begin{centering}
\includegraphics[width=0.4\textwidth]{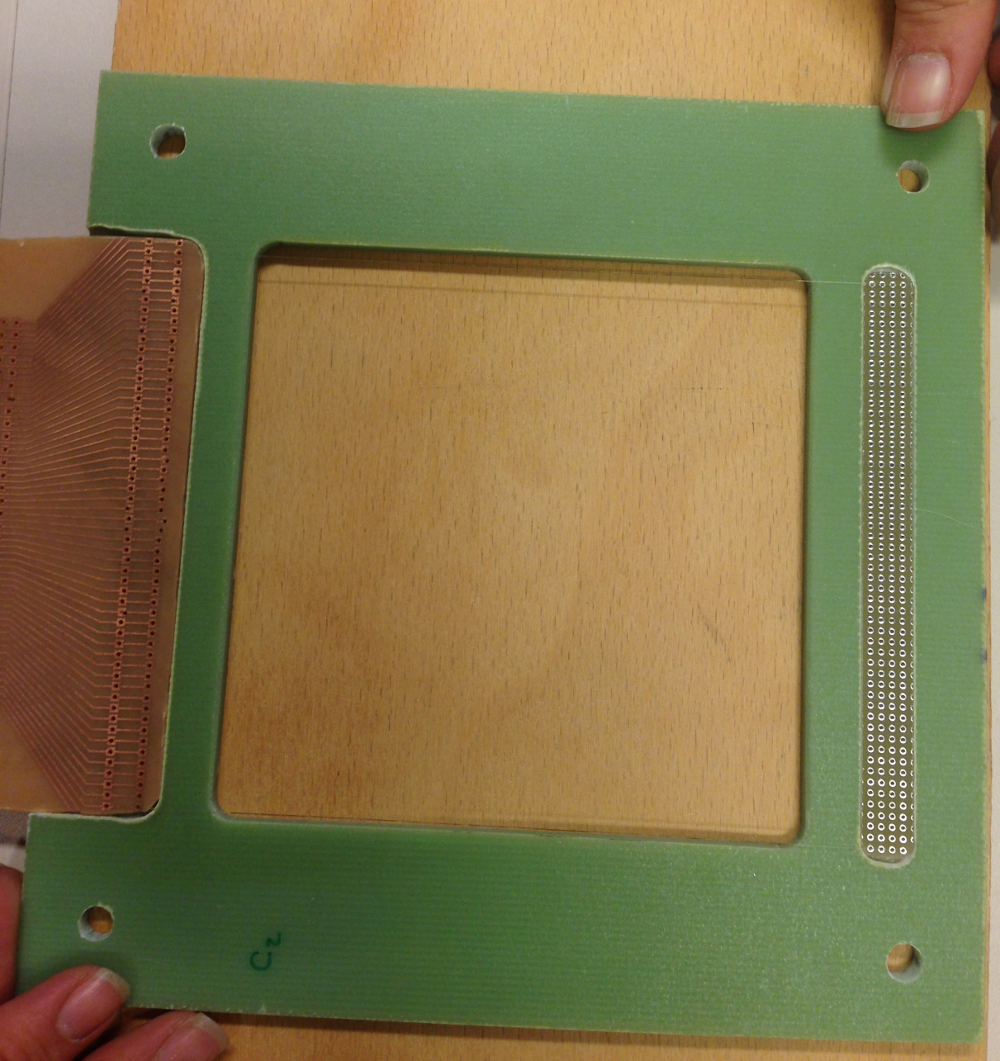}
\par\end{centering}

\protect\caption{Installing the wires\label{fig:wire-test}}
\end{figure}

After completion of the frames, the next task is the attachment of
various components such as the gas input output connectors and wire
holding PCBs. These pieces are glued to the FR4 material using epoxy
paste which would not contaminate the ionization gas. Epoxy paste
is strong enough to keep the wires in tension when the PCBs are glued
and it can be found in local hardware stores or in big markets. Instead
of using a machine to ensure the same mechanical tension across the
conductive wires, a simple method is used: one side of the wire is
connected to the PCB and the other side is connected to a fishing
weight of about 100gr. These weights can be found in local sports
store and can be pre-adjusted to the same mass with a simple kitchen
scale, which is usually accurate to a gram. Fig. \ref{fig:wire-test}
contains the second cathode plane with two wires installed on each
side. Note that the wire pitch converter is on the left hand side.
At the outer part of the pitch converter, a standard 60 pin connector
is soldered. 

After installing the wires on the PCBs, the thin film should be glued
to the upper and lower FR4 layers. For this purpose Kapton film with
25 micron thickness is selected, as it is durable under pressure and
stable under temperature changes \cite{kapton}. Instead of Kapton,
one can also use Mylar which may be easier to procure. However, Kapton
is preferred for its even thinner sheet availability (down to 13 microns),
which would allow even less energetic particles to enter the chamber.
Fluka simulations show that protons of 1.3 MeV can travel about 30
microns inside the Kapton film, while for 1.4 MeV protons the path
length increases to 33 microns \cite{Fluka-a,Fluka-b}. For gas tightness,
o-rings of 5mm width and about 1mm thickness can be used. To drive
down the cost of the detector, it is envisaged to produce this square
o-ring from the packaging rubbers of appropriate shape and size. In
this project, 20cm long big rubber bands are cut to be fit in the
o-ring ditches and glued at each end with a strong glue, forming a
complete square of side 16cm. After fitting the o-rings, all the layers
are stacked consecutively allowing the cathode wires to be parallel
and the anode wires perpendicular in between. Then screws of 6mm are
sticked at four corners and tightened.

\section{Detector Tests}

First item to be tested is gas tightness since a certain level of
gas pureness needs to be achieved. One also needs to be careful about
the materials as some of the materials like the glue may easily cause
contamination in the gas. If the gas in the detector is not pure enough,
then when the high voltage is applied, the wires might produce sparks
which in turn, may cause harm to the detector or the electronic circuitry.
To test gas tightness, when all the individual construction work is
finished but before installing the wires, the assembled chamber is
filled with gas. By applying a leak detector spray, possible leaks
are examined. Leak detection can also be done with simple soap foam
but then it should be cleaned very carefully. In case of any leak,
one can change the o-rings or apply more epoxy paste locally. Only
after being sure that the detector is gas tight, and the installation
of the wires is complete, the high voltage can be applied by increasing
it slowly and observing the detector for any sparks. After the gas
tightness and the high voltage tests, the next test is for electronic
circuitry. By sending analog signals from a signal generator to a
channel in the readout system, it is possible to observe the final
signal from the oscilloscope. The final two signals should be compatible
with the expected values depending on the channel and the delay step
of the microchips used. At this point the detector becomes ready for
particles, in particular cosmic particles (muons) for the first tests.
The cosmic muons should be equally distributed on the DWC active area.
Further studies could be made with the new detector installed between
two previously calibrated and tested DWCs. Again testing with cosmic
muons, one should be able to find the track of the cosmic particles
using the older DWCs and check the efficiency of the new detector
by comparing its hit position to the fitted track.

\section{Conclusions and outlook}

Although this particular detector is being produced as a close copy
of the DWCs in operation at CERN, a number of investigations possibly
leading to improvements are being done. For example the geometrical
effects, high voltage settings, wire material and pitch as well as
the gas mixture ratios can be investigated by the means of the well
established gaseous detector simulation tool, Garfield \cite{garfield}.
Additionally this chamber can be operated in fully gas tight mode,
i.e. without the gas flow. This would be extremely useful for educational
and demonstration purposes where the typical operation time is few
hours at most. Such a demonstration can be augmented by a small portable
computer running an oscilloscope and display application. There are
such low cost USB devices \cite{bitscope}and computers capable of
running these \cite{raspberry}, however the delay microchips would
need to be adapted to the low resolution of the scopes. With such
a setup and a self triggering DWC, it would be possible to display
the cosmic rays in real-time. Furthermore, if batteries can be used
as both low voltage and high voltage power sources, the detector would
become cable-independent, a major step for using it in particle physics
education.
\begin{acknowledgments}
The authors would like to thank M. Joos and J. Spanggaard for useful
discussions and suggestions, TAEK machine shop for the help on PCB
machining and S. Sekmen for a careful reading of the manuscript.\end{acknowledgments}

\end{document}